\documentclass[fleqn,10pt]{wlscirep}
\usepackage[utf8]{inputenc}
\usepackage[T1]{fontenc}
\usepackage{physics}
\usepackage{color,xcolor}
\usepackage{cite}
\usepackage{physics}
\usepackage{array}
\usepackage{multirow}
\usepackage{bigstrut}
\usepackage{longtable}

\title{A universal simulating framework for quantum key distribution systems}

\author[1,2,3]{Guan-Jie FAN-YUAN}
\author[1,2,3,*]{Wei CHEN}
\author[1,2,3]{Feng-Yu LU}
\author[1,2,3]{Zhen-Qiang YIN}
\author[1,2,3]{\\Shuang WANG}
\author[1,2,3]{Guang-Can GUO}
\author[1,2,3,*]{Zheng-Fu HAN}

\affil[1]{CAS Key Laboratory of Quantum Information, University of Science and Technology of China, Hefei, Anhui 230026, China}
\affil[2]{CAS Center for Excellence in Quantum Information and Quantum Physics, University of Science and Technology of China, Hefei, Anhui 230026, China}
\affil[3]{State Key Laboratory of Cryptology, P. O. Box 5159, Beijing 100878, China}
\affil[*]{Corresponding author: weich@ustc.edu.cn, zfhan@ustc.edu.cn}

\begin{abstract}
Quantum key distribution (QKD) provides a physical-based way to conciliate keys between remote users securely. Simulation is an essential method for designing and optimizing QKD systems. We develop a universal simulation framework based on quantum operator descriptions of photon signals and optical devices. The optical devices can be freely combined and driven by the photon excitation events, which make it appropriate for arbitrary QKD systems in principle. Our framework focuses on realistic characters of optical devices and system structures. The imperfections of the devices and the non-local properties of a quantum system are taken into account when modeling. We simulate the single-photon and Hong–Ou–Mandel interference optical units, which are fundamental of QKD systems. The results using this event-driven framework agree well with the theoretical results, which indicate its feasibility for QKD.
\end{abstract}

\begin{document}
\flushbottom
\maketitle
\thispagestyle{empty}

\maketitle

\section{Introduction}
Quantum key distribution (QKD)\cite{bennet1984quantum, ekert1991quantum} can generate keys between remoter users in public channels against the threat of quantum computing\cite{gottesman2004security, scarani2009security}. Since its first protocol proposed\cite{bennet1984quantum} in 1984, QKD has achieved significant progress in theory\cite{wang2005beating,lo2005decoy,tomamichel2011uncertainty,laing2010reference,yin2014reference,lo2012measurement,curty2014finite,sasaki2014practical,lim2014concise,rusca2018finite,lucamarini2018overcoming,ma2018phase,wang2018twin,cui2019twin} and experiment\cite{Peng2007Experimental,Dixon2008Gigahertz,Wang2015Experimental,Wang2015Phase,takesue2015experimental,Yin2016Measurement,Comandar2016Quantum,liao2017satellite,Frohlich2017Long,boaron2018secure,Wang2018Practical,wang2019beating,Minder2019Experimental,Zhong2019Proof,Liu2019Experimental,Chen2019Sending}. However, there are still some challenges for QKD on its road to real-life applications\cite{Lo2014Secure,Scarani2009The,yoshino2018quantum,angrisani2008modeling,wang2016non-markovian,zhang2015advances,fan2018afterpulse}. One of the essential challenges is the gap between the practical systems and the theoretical model of QKD \cite{Lo2014Secure,Scarani2009The} since the device imperfections are in multiple dimensions and hardly to theoretically evaluate, especially in a comprehended system. Fortunately, a reliable simulation model will significantly benefit the design and analysis of an optical communication system\cite{buhari2012efficient,mailloux2015modeling,archana2015implementation} with no exception of QKD.

In general, there are two kinds of simulation models aiming at different targets. Most of the existing QKD simulations are focusing on the theoretical part of the protocols, and always try to catch a tight bound of the secure key rate(SKR)\cite{lo2005decoy,ma2005practical,tomamichel2011uncertainty,curty2014finite,lim2014concise,rusca2018finite}. The realization scheme of the protocols, as well as the imperfection of the devices, are essential in practical QKD security evaluations. Unfortunately, these vital factors have not been considered meticulously in most of the existing simulation works. From the perspective of signals and systems, a quantum system can be regarded as the transformation of quantum signals, and the other QKD simulations are model-based design (MBD)\cite{buhari2012efficient,mailloux2015modeling,archana2015implementation,krenn2016automated}. The components in MBD-based frameworks are modeled individually and can build complex systems by combining these devices.

Furthermore, with the help of searching and optimizing algorithms, MBD simulation frameworks will benefit from designing and evaluating quantum information processing systems\cite{krenn2016automated}. Regrettably, the major flaw in existing MBD-based models for QKD is the quantum states and devices are depicted using classical electromagnetic field (EMF) theory. Therefore, it is difficult to precisely simulate some specific quantum procedures in the quantum field, for example, the sub-Poissonian photon statistics and anti-bunching of photons.

QKD is a procedure involving the preparation, propagation, transformation, and measurement of quantum states. In order to subtly evaluate the practical QKD systems, as well as their imperfections, we developed a universal framework for QKD modeling with quantum descriptions. We describe the quantum states of photons and optical components in quantum operators. The states and the devices have multiple dimensions and parameters which can be adjusted independently. Thus the practical characteristics of the optical devices and the disturbance from the eavesdropper or the environment can be simulated comprehensively. We employ c++ language and combined our self-designed packages into OMNeT++\cite{varga1999using,varga2001discrete,varga2008overview}, which is widely adopted in simulations of classical optical systems and networks. The devices are built independently and can be combined to build a complex QKD system. With the software package, we successfully verify single-photon and Hong–Ou–Mandel interference\cite{hong1987measurement} optical units, which are the kernel of the BB84\cite{bennet1984quantum} and measurement-device-independent \cite{lo2012measurement} QKD protocols. The simulation results indicate that this quantum-described framework is available for QKD simulation. Furthermore, since OMNet++ are employed widely in simulations of classical optics communication networks, this work shows the feasibility of extending the classical simulation platform like OMNet++ into the field of quantum research with quantum descriptions.

Firstly, we show the design methods of the simulation framework in section 2. In section3, we describe the models of the optical elements in QKD systems, which are the quantum light source and the photons states, the transformation of the photon states with optical devices, and the function of SPDs. Section 4.gives the simulation results of the single-photon and Hong–Ou–Mandel (HOM) interference with our framework. Finally, we give a conclusion and a short discussion.

\section{Design of the simulation framework}

\begin{figure}[!htbp]
	\centering\includegraphics[width=0.86\textwidth]{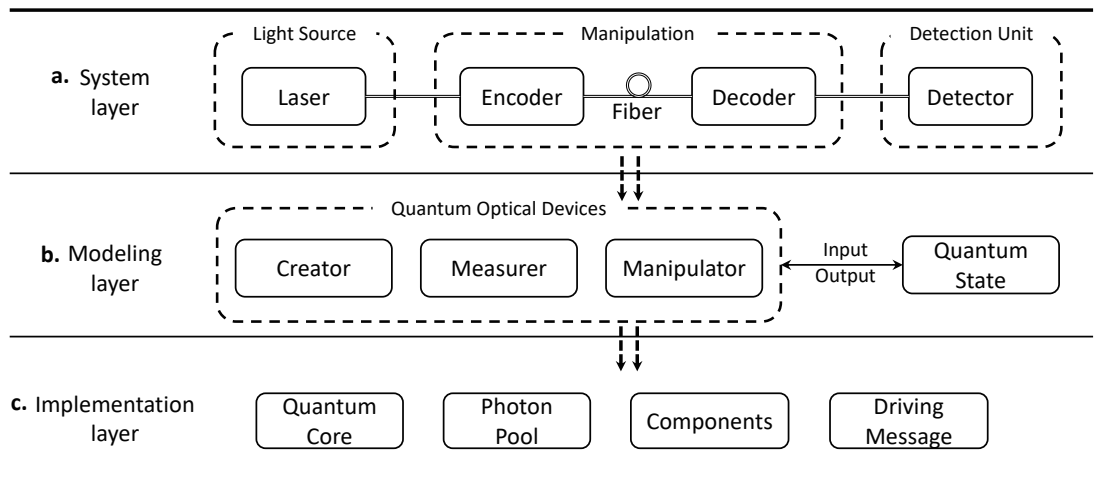}
	\caption{Model construction. \textbf{a.} system layer: the tripartite optical devices of real-world QKD systems, which consist of the light source, manipulation, and detection units. The quantum states of photons are created by the light source, manipulated by the systems and the channels, and finally measured by detectors. \textbf{b.} Modeling layer: devices and procedure modeling. Quantum optical devices are modeled as creator, measurer, and manipulator according to the three types of quantum operations in \textbf{a}. Quantum states transfer the optical devices and are transformed.  The quantum signals are passed through the devices through their input and output ports. \textbf{c.} Implementation layer: Realization of the models in modeling layer using C++ classes integrated with OMNet++. Specifically, each optical device is encapsulated as a component class, of which the input and output ports are driven by messages. The quantum pool is a custom-designed array to store the same kind of quantum states, and the quantum core is the processor of the quantum operations like Pauli operators and the projection measurements.} 
	\label{ModelStructure}
\end{figure}

A mature QKD modeling framework should achieve equilibrium of multi-features, such as accuracy, scalability, efficiency, operability, compatibility, and cost. Our prime motivation is to design an elaborate modeling framework for the signals, elements, and the procedure in a quantum system with quantum representations. Therefore, we pay more attention to the simulation accuracy of the framework and focus on a discrete-variable-based QKD systems in this work. We design the simulation framework in three layers: the system layer, modeling layer, and implementation layer. The construction and realization of this simulation framework are shown in Fig.\ref{ModelStructure}. 

Photons are natural carriers for quantum information. The optical units in a QKD system generally execute three steps, which are the preparation of photon states with the light source, manipulation of photon states using optical components, and measurement of photons by detectors. The system layer maps the real-life QKD systems into these three types of simulation units, handles their connections, and dispose of the interaction between the users and the abstract underlying data. 

The modeling layer takes charge of the abstract units and their data structures. There are two types of data structures, which are quantum optical devices and quantum signals. The former described the behaviors of different types of optical devices when photons arrived, and the latter contains the quantum states entering and exiting of the optical devices. Besides providing the quantum description of signals and components, we use the discrete-event driving scheme to handle the quantum processing procedure of the system. The modeling layer is the kernel of the simulation framework, and the details are described in Section \ref{sec:model}.

The implementation layer contains three major packages to handle the evolution of quantum states (QuantumCore), to storage the homogeneous quantum states (PhotonPool), and to process the driving message of arriving photons (Components). We use C++ language to implement the data structure, handle the messages, and execute the procedures (also described in section 3). We integrate these C++ packages into OMNet++ platform\cite{varga1999using,varga2001discrete,varga2008overview}, and use its visual interface and message-driving engine to verify the simulation. 

\section{\label{sec:model}Modeling of quantum state and optical components}

The process of QKD can be regarded as the transformations of the quantum states by optical components. Thus, the descriptions of the quantum states are foundational to make the simulation model universal and expansible. The method to descript the quantum states is shown in Section \ref{subsec:qs}. Furthermore, in Section \ref{subsec:ocp}-\ref{subsec:ocd},  we introduce the modeling of optical components in the order of sequential operations of the quantum states (preparation, manipulation, and measurement).

\subsection{\label{subsec:qs}Modeling of quantum state}

A typical light source for QKD, no matter the weak coherent-state source, sub-Poissonian source, or single-photon source, can be represented with the basis of Fock states in principle. For example, phase-randomized coherent-state light source\cite{glauber1963coherent,lo2005decoy}, which has been widely adopted in practical QKD systems, can be described as
\begin{equation}
\rho=\int_0^{2\pi}\frac{\dd\theta}{2\pi}\ket{\alpha}\bra{\alpha}=\sum_{n}\frac{e^{-\mu}\mu^n}{n!}\ket{n}\bra{n}
\label{eq1}
\end{equation}
where $\mu=|\alpha^2|$ is the average photon number, while the single-photon state can be regarded as a crucial point of Fock states.

In order to simulate practical QKD systems, the descriptions of the photon states should contain frequently-used physical degrees of freedom. At present, our framework includes five independently operable degrees of freedom for QKD encoding, which are time, path (momenta), phase, polarization, and frequency. The photon state created in path $\alpha$ is given by
\begin{equation}
\ket{\psi}_\alpha=C\int\dd\omega\phi(\omega)e^{-i(\omega\tau -\varphi)}(\alpha\vu*{a}_H^\dagger(\omega)+\beta e^{-i\theta}\vu*{a}_V^\dagger(\omega))\ket{0}
\label{eq2}
\end{equation}
where $C$ is the normalization coefficient, $\phi(\omega)$ is the frequency distribution assumed to be the Gaussian profile $\phi(\omega)=\frac{e^{-(\omega-\omega_\mu)^2/4\sigma^2}}{(2\pi)^{1/4}\sqrt{\sigma}}$, $\vu*{a}_H^\dagger(\omega)$ represents a creation operators acting on a single frequency mode $\omega$ and a polarization mode $H$, $\vu*{a}_V^\dagger(\omega)$ is similar. $\alpha$ and $\beta e^{-i\theta}$ are the components of Jones vector\cite{jones1941new}. The normalization requires that $\int\dd\omega{|\phi(\omega)|}^2=1$ and ${|\alpha|}^2+{|\beta e^{-i\theta}|}^2=1$. 

We create a photon state class with 9 independently tunable variables according to Eq.\ref{eq2}, as shown in Table \ref{tab1}. The parameters of spectrum (SpectralMu and SpectralSigma), pulse width (Delay) and polarization (Alpha, Beta, DeltaPhase) base on the characteristics of the light source user employed. RouteID depends on the connections of the optical devices. $C$ is initialized by $\left(\frac{1}{\sqrt{n!}}\right)^\frac{1}{n}$ to ensure the normalization of $n$-photon Fock state, $\frac{\left(\vu*{a}^\dagger\right)^n}{\sqrt{n!}}\ket{0}$.

\begin{table}[!htbp]
	\footnotesize
	\caption{Member Variables Definition of Photon State}
	\label{tab1}
	\tabcolsep 10pt 
	\begin{tabular*}{\textwidth}{ccp{300pt}}
		\toprule
		Variable Name		&Data Type		&Explanation \\\hline
		Delay					&double			&The relative time delay between the photon states which in a same multi-photon states system\\
		RouteID				&double			&The path where the photon state is propagating\\
		SpectralMu			&double			&The mean of the Gaussian frequency distribution\\
		SpectralSigma			&double			&The standard deviation of the Gaussian frequency distribution\\
		Phase					&double			&The relative phrase between the photon states which in a same multi-photon states system\\
		Alpha					&double			&The amplitude of the field in the horizontal direction\\
		Beta					&double			&The amplitude of the field in the vertical direction\\
		DeltaPhase			&double			&The difference between the phase angles of fields in horizontal and vertical directions\\
		Coefficient				&double			&The normalization coefficient\\
		\bottomrule
	\end{tabular*}
\end{table}

\begin{figure}[!htbp]
	\centering\includegraphics[width=0.7\textwidth]{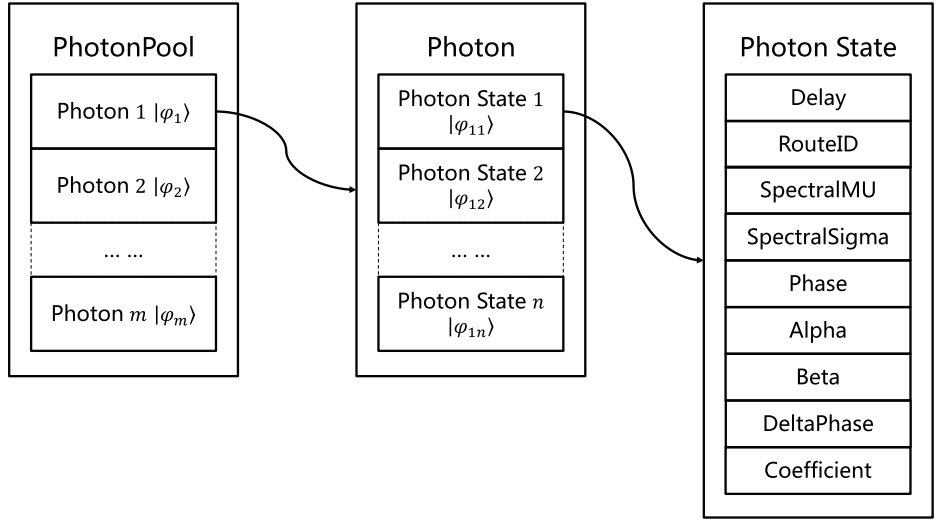}
	\caption{Data structure for the depiction of PhotonPool. The quantum state, as a constitution of photons, is stored in a PhotonPool which is an array of photons. In PhotonPool, each photon consists of the superposition of photon states, which is the basic element of the data structure and contains the information of the degrees of freedom listed in Table \ref{tab1}.}
\label{DataStructure}
\end{figure}

According to the design above, we can depict the quantum states of the photons with the basis of Fock states. Furthermore, we merge the photons from an individual system with common attributes, for example, the photons from a laser source into a common data set. This data set noted as PhotonPool, which is dynamic in our modeling framework, can be created by emitting a light pulse from a laser source, merged by the correlation among individual systems, and pruned by measuring all its photon. In a PhotonPool, each photon can be described with the superposition of the photon states mentioned above, and the photon state is the fundamental element of this data structure. The data structure is shown in Fig. \ref{DataStructure} and the quantum state of a PhotonPool can be described as

\begin{equation}
\ket{\psi}_\text{PhotonPool}=\left(\ket{\psi}_{11}+\ket{\psi}_{12}+...\right)\left(\ket{\psi}_{21}+\ket{\psi}_{22}+...\right)\times...=\prod_{m}\sum_{n}\ket{\psi}_{mn}
\end{equation}
where $\ket{\psi}_{mn}$ denotes the $n^{th}$ photon state of the $m^{th}$ photon.

\subsection{\label{subsec:ocp}Modeling of the quantum-state preparation}
In our model, the pulsed laser prepares a quantum state is equivalent to create and initialize a PhotonPool. The photon number of an initialized PhotonPool is a random number generated by a random number generator with a specific probability distribution related to the character of the laser source. For example, for a weak-coherent source, the photon number follows a Poisson distribution, as shown in Eq. (\ref{eq1}).

In addition, each photon of the PhotonPool initially contains only one photon state without any manipulation. The initial value of the variables listed in Table \ref{tab1} is assigned according to the character of a laser source. For undefined characters, the initial values can be constants or random numbers with a uniform distribution, according to the requirement of the users.

Therefore, the quantum state of an initialized PhotonPool can be expressed as
\begin{equation}
\ket{\psi}_\text{NewPhotonPool}=\ket{\psi}_{11}\ket{\psi}_{21}...=\prod_{m}\ket{\psi}_{m1}
\end{equation}
where the $\ket{\psi}_{m1}$ is an initialized photon state of $m^{th}$ photon. $\ket{\psi}_{m1}$ can be derived by Eq. (\ref{eq2})
\begin{equation}
\ket{\psi}_{m1}=C_0\int\dd\omega\phi(\omega)e^{-i(\omega\tau_0 -\varphi_0)}(\alpha_0\vu*{a}_H^\dagger(\omega)+\beta_0e^{-i\theta_0}\vu*{a}_V^\dagger(\omega))\ket{0}
\end{equation}
where the subscripts represent the initial value.

Moreover, the size of PhotonPool is finite, however, a phase-randomized coherent state has infinite components of Fock states. Therefore, a truncation is essential to the practical simulation. The users can select the starting Fock state of the truncation on demand. Also, here we provide a reference method of truncation.

Because the photon number of a phase-randomized coherent state is a random variable following the Poisson distribution and the number of the trials, $N_\mu$, that preparing the weak coherent state with a mean photon number, $\mu$, in a QKD task is finite, the truncation point, $n_t$, can be given by the minimal solution which satisfies the following equation,
\begin{equation}
\epsilon_t=1-(\sum_{n=0}^{n_t}\frac{e^{-\mu}\mu^n}{n!})^{N_\mu}
\end{equation}
where $\epsilon_t$ is a security parameter, which indicates the probability that the maximal photon number in $N_\mu$ trials is not less than $n_t$. So the photon numbers which are greater than $n_t$ can be discarded with a failure probability of $\epsilon_t$.

\subsection{\label{subsec:ocm}Modeling of the quantum-state manipulation}
The photon states can be created, modified, or annihilated by optical components. The linear optical devices in QKD systems, as well as the quantum channels like fiber and free-space, are regarded as the manipulations of the photon states. These devices can be divided into two categories according to whether the device contains path-depending operation. The path-depending operation completed by a beamsplitter (BS) or a  polarization beamsplitter (PBS), and other optical elements only change the parameters of the photon states without creating or merging any optical path.

When a BS or a PBS splits the path states of the photons from one path to two paths, new photon states with different values of RouteID are created and added to the state set of the photons. The BS and PBS also can correlate the incident photons from different paths. Therefore, the merging of PhotonPools can occur if the incident photons belonging to different PhotonPool. 

The path-splitting operations of BS and PBS can be described as following
\begin{equation}
\vu*{a}^\dagger(\omega)\rightarrow\sqrt{T}\vu*{c}^\dagger(\omega)-\sqrt{R}\vu*{d}^\dagger(\omega)
\label{eq4}
\end{equation}
\begin{equation}
\vu*{b}^\dagger(\omega)\rightarrow\sqrt{T}\vu*{c}^\dagger(\omega)+\sqrt{R}\vu*{d}^\dagger(\omega)
\label{eq5}
\end{equation}
\begin{equation}
\vu*{a}_H^\dagger(\omega)\rightarrow\vu*{c}_H^\dagger(\omega), \vu*{a}_V^\dagger(\omega)\rightarrow\vu*{d}_V^\dagger(\omega)
\label{eq6}
\end{equation}
\begin{equation}
\vu*{b}_H^\dagger(\omega)\rightarrow\vu*{d}_H^\dagger(\omega), \vu*{b}_V^\dagger(\omega)\rightarrow\vu*{c}_V^\dagger(\omega)
\label{eq7}
\end{equation}
where T and R are the transmissivity and the reflectivity of the elements, respectively. Eq.\ref{eq4}-\ref{eq5} and Eq.\ref{eq6}-\ref{eq7} describe the actions of a BS and a PBS, individually.

Eq. \ref{eq4}-\ref{eq7} characterize the behaviors of the BS and PBS. However, the parameters like transmission loss and the extinction ratio of the optical devices should match the parameters of the off-the-shelf devices. The modified equations are
\begin{equation}
\vu*{a}^\dagger(\omega)\rightarrow\sqrt{10^\frac{-l}{10}T}\vu*{c}^\dagger(\omega)-\sqrt{10^\frac{-l}{10}R}\vu*{d}^\dagger(\omega)
\label{eq9}
\end{equation}
\begin{equation}
\vu*{b}^\dagger(\omega)\rightarrow\sqrt{10^\frac{-l}{10}T}\vu*{c}^\dagger(\omega)+\sqrt{10^\frac{-l}{10}R}\vu*{d}^\dagger(\omega)
\label{eq10}
\end{equation}
\begin{equation}
\vu*{a}_H^\dagger(\omega)\rightarrow\sqrt{10^\frac{-l_H}{10}\frac{R_E}{R_E+1}}\vu*{c}_H^\dagger(\omega), 
\vu*{a}_V^\dagger(\omega)\rightarrow\sqrt{10^\frac{-l_V}{10}\frac{R_E}{R_E+1}}\vu*{d}_V^\dagger(\omega)
\label{eq11}
\end{equation}
\begin{equation}
\vu*{b}_H^\dagger(\omega)\rightarrow\sqrt{10^\frac{-l_H}{10}\frac{R_E}{R_E+1}}\vu*{d}_H^\dagger(\omega), 
\vu*{b}_V^\dagger(\omega)\rightarrow\sqrt{10^\frac{-l_V}{10}\frac{R_E}{R_E+1}}\vu*{c}_V^\dagger(\omega)
\label{eq12}
\end{equation}
where $R_E$ is the extinction ratio ,$l$ is the loss, $l_H$ and $l_V$ are the polarization dependent loss (PDL). Eq.\ref{eq9}-\ref{eq10} are for a BS and Eq.\ref{eq11}-\ref{eq12} are for a PBS.

According to Eq.\ref{eq9}-\ref{eq12}, the tunable parameters of BS and PBS are shown in Table \ref{tab2}, which can be customized adjusted according to the specific components and their measurement results in practical experiments.

\begin{table}[ht]
\footnotesize
\caption{Member Variables Definition of BS and PBS}
\label{tab2}
\tabcolsep 8pt 
\begin{tabular*}{\textwidth}{cccc}
\toprule
  Component&Variable Name&Data Type&Explanation\\\hline
Beamsplitter&SplittingRatioR&double&Reflectance\\
&SplittingRatioT&double&Transmittance\\
&Loss&double&Loss\\
&ExtinctionRatio&double&ExtinctionRatio\\\hline
Polarization Beamsplitter&Loss&double&Loss\\
&LossH&double&The loss of the field in the horizontal direction\\
&LossV&double&The loss of the field in the vertical direction\\
\bottomrule
\end{tabular*}
\end{table}

Optical devices without path state operation are listed in \ref{tab3} and shown in Appendix A in detail. Each optical device is an individual component in the optical device library of the simulation framework and can be instantiated to multiple units and combined to build a simulation scheme. The final states of the incoming photons that go through the whole system can be obtained by calculating the transformations of these units. 

Moreover, the channel disturbance is included in the design of the fiber component. Specifically, the random walk theory is used to simulate the normal distributed random disturbance of fiber to the parameters of photon states.
\begin{equation}
a_o=a_i+\delta
\end{equation}
where $a\in\{\varphi, \alpha, \beta, \theta\}$ is one of the parameters of photon state, the subscripts $o$ and $i$ represent the output and input, respectively. $\delta$ is the random disturbance that follows a normal distribution,
\begin{equation}
\delta\sim N(\mu_a,\sigma_a)
\end{equation}
where $\mu_a$ and $\sigma_a$ is the expectation and variance of the distribution and can be estimated by practical data.

The fiber component is common for the polarization maintaining fiber and the single mode fiber by modulating the polarization-dependent $\mu$ and $\sigma$.

\begin{table}[ht]
\footnotesize
\caption{Common modeled optical elements}
\label{tab3}
\tabcolsep 20pt 
\begin{tabular*}{\textwidth}{cccc}
\toprule
  Attenuator&Bandpass Filter&Circulator&Polarization Modulator\\
  Phase Modulator&Isolator&1x2 Optical Switch&Waveplate\\
  Fiber&Faraday Mirror (FM)&&\\
\bottomrule
\end{tabular*}
\end{table}

\subsection{\label{subsec:ocd}Modeling of the single-photon detector}
In discrete-variable QKD systems, the photon state propagating in a specific path arrived in a single photon detector (SPD) module after creations and transformations. The measurement of photon states are fulfilled by the SPDs together with the optical elements for QKD decoding, and then the photon annihilates. 

The simulating system processes the response of SPDs according to the information of the photons stored in PhotonPool. Firstly, the SPD module sends the necessary information to a processing unit named QuantumCore, including the ID of measuring PhotonPool, and the RouteID represented the path where the SPD is located. Secondly, QuantumCore calculates the photon number arrived at SPD and sends the reserved quantum states back to the QuantumPool according to the models of the quantum states and the optical devices mentioned above. Finally, the SPD module calculates the click probability based on its parameters and the photon number that arrived at the SPD. The detailed procedure is explained in the next section. 

\subsubsection{Calculation of the photon numbers}
There are two critical calculations, which are the photon number after projection measurements and the click probability of the photons. QuantumCore firstly calculates the arriving probabilities of the different number of photons according to the final quantum states exited from the optical devices. The probability of 0...n photons arrived at the SPD after projection measurement is $p_0, p_1, ... , p_n \left(\sum_{n}p_n=1\right)$, where n is the photon number with the same path ID as the SPD, which is also the maximum photon number which can arrive at the SPD. Following the Monte Carlo method\cite{metropolis1949monte}, we randomly assign the instant light pulse arriving at the SPD in a specific path as the Fock states with photon number m ($0<=m<=n$) according to the probabilities of $p_0$ to $p_n$. As a demonstration, we calculate $p_n$ in a HOM interference with ideal single photons\cite{hong1987measurement}.
\begin{figure}[ht!]
	\centering\includegraphics[width=0.5\textwidth]{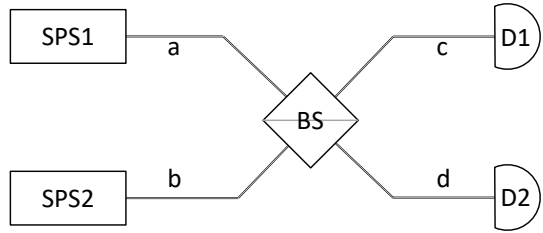}
	\caption{Hong-Ou-Mandel interference at a beam splitter. SPS, single-photon source; BS, beamsplitter; D, detector.}
\end{figure}

The two-photon input state is described as
\begin{equation}
\begin{split}
\ket{\psi}_{in}=&C_1\int\dd\omega_1\phi(\omega_1)e^{-i(\omega_1\tau_1 -\varphi_1)}(\alpha_1\vu*{a}_H^\dagger(\omega_1)+\beta_1e^{-i\theta_1}\vu*{a}_V^\dagger(\omega_1))\\
&\times C_2\int\dd\omega_2\phi(\omega_2)e^{-i(\omega_2\tau_2 -\varphi_2)}(\alpha_2\vu*{a}_H^\dagger(\omega_2)+\beta_2e^{-i\theta_2}\vu*{a}_V^\dagger(\omega_2))\\
=&\left(\vu*{a}_{1H}^\dagger+\vu*{a}_{1V}^\dagger\right)\left(\vu*{b}_{2H}^\dagger+\vu*{b}_{2V}^\dagger\right)\ket{0}
\end{split}
\end{equation}
where
\begin{equation}
\begin{split}
\vu*{a}_{H}^\dagger=C\int\dd\omega\phi(\omega)e^{-i(\omega\tau -\varphi)}\alpha\vu*{a}_H^\dagger(\omega)\\
\vu*{a}_{V}^\dagger=C\int\dd\omega\phi(\omega)e^{-i(\omega\tau -\varphi)}\beta e^{-i\theta}\vu*{a}_V^\dagger(\omega)
\end{split}
\end{equation}

The action of BS is given by Eq.\ref{eq9}-\ref{eq12}, and the output state is
\begin{equation}
\ket{\psi}_{out}=\left(\vu*{c}_{1H}^\dagger-\vu*{d}_{1H}^\dagger+\vu*{c}_{1V}^\dagger-\vu*{d}_{1V}^\dagger\right)\left(\vu*{c}_{2H}^\dagger+\vu*{d}_{2H}^\dagger+\vu*{c}_{2V}^\dagger+\vu*{d}_{2V}^\dagger\right)\ket{0}
\end{equation}

When the measurement occurs in path $c$, $\ket{\psi}_{out}$ indicates three possible numbers of photon
\begin{equation}
\begin{split}
&\ket{\psi_0}=\left(-\vu*{d}_{1H}^\dagger-\vu*{d}_{1V}^\dagger\right)\left(\vu*{d}_{2H}^\dagger+\vu*{d}_{2V}^\dagger\right)\ket{0}\\
&\ket{\psi_1}=\left(\left(\vu*{c}_{1H}^\dagger+\vu*{c}_{1V}^\dagger\right)\left(\vu*{d}_{2H}^\dagger+\vu*{d}_{2V}^\dagger\right)\left(-\vu*{d}_{1H}^\dagger-\vu*{d}_{1V}^\dagger\right)\left(\vu*{c}_{2H}^\dagger+\vu*{c}_{2V}^\dagger\right)\right)\ket{0}\\
&\ket{\psi_2}=\left(\vu*{c}_{1H}^\dagger+\vu*{c}_{1V}^\dagger\right)\left(\vu*{c}_{2H}^\dagger+\vu*{c}_{2V}^\dagger\right)\ket{0}
\end{split}
\end{equation}
Thus, the probabilities of different photon numbers are given by $p_{i}=\bra{\psi_i}\ket{\psi_i}$, where $i$=0, 1, and 2. For example, for the case of $p_0$
\begin{equation}
\begin{split}
p_{0}&=\bra{\psi_0}\ket{\psi_0}\\
&=\bra{0}\left(-\vu*{d}_{1H}-\vu*{d}_{1V}\right)\left(\vu*{d}_{2H}+\vu*{d}_{2V}\right)\left(-\vu*{d}_{1H}^\dagger-\vu*{d}_{1V}^\dagger\right)\left(\vu*{d}_{2H}^\dagger+\vu*{d}_{2V}^\dagger\right)\ket{0}
\end{split}
\label{eq18}
\end{equation}

By such equations, we obtain the probability distribution of the Fock states with different photon numbers. QuantumCore returned the photon number and updated the variables of the measured PhotonPool for the next measurement. Since the detection of SPD is essentially the projection of the quantum states in a unique path where the SPD lays, and the projection measurement will collapse a photon into a specific state. Therefore, the photons projected successfully would be removed from the PhotonPool, and all states of remaining photons with the identical RouteID of the SPD but did not be detected are removed simultaneously. The normalization coefficient of the remaining photon states would be recalculated and updated. The removal of projected photons completes the local measurement on a path. In addition, the update of the remaining photon correlates this measurement with other measurements of new PhotonPool and reflects the non-local properties of a quantum system.

\subsubsection{Calculating the clicking probability}
The clicking events of a SPD are generally composed of three parts, which are the photons arrived, the dark counts (dark current), and the after pulses. We build the SPD module with seven elements, including the response of the photons and their electrical parameters, to cover the realistic characters of the SPD.

\begin{table}[ht]
\footnotesize
\caption{Member Variables Definition of SPD}
\label{tabSPD}
\tabcolsep 10pt
\begin{tabular*}{\textwidth}{ccp{260pt}}
\toprule
		Variable Name			&Data Type		&Explanation\\\hline
		DetectionEfficiency		&double			&The detection efficiency of the given wavelength\\
		ProbabilityDarkCount		&double			&The dark count probability of the SPD\\
		ProbabilityAfterpulse		&double			&The afterpulse probability of the SPD\\
		TimingJitter				&double			&The jitter of click signal emission time\\
		ResolvesPhotonNumber	&bool				&The flag of photon number resolution, when it is true, the SPD return photon number, otherwise it return click signal\\
		Enable						&bool				&The flag of SPD on-off state, when it is true, the SPD responds the photon pulse, otherwise it does not work\\
		TimeWidth				&double			&The duration of open gate for gate mode\\
\bottomrule
\end{tabular*}
\end{table}

The parameters and their definitions are listed in Table \ref{tabSPD}. The clicking probability $p_c$ can be calculated using the formula
\begin{equation}
p_c=1-((1-\eta)^n(1-p_d\delta t)(1-p_a))
\end{equation}
where $\eta$ is the detection efficiency, $n$ is the photon number, $p_d$ is the dark count probability, $\delta t$ is the duration of gating time and $p_a$ is the afterpulse probability of SPD\cite{fan2018afterpulse}.

\section{Simulation Results}

The QKD session is the statistical result of plenty of independent photons with the procedure of preparation, manipulation, and measurement. The encoder and the decoder (codec) is the kernel to perform QKD protocols. Therefore, we simulate the Mach-Zehnder interferometer (MZI)\cite{zehnder1891neuer,mach1892ueber} and HOM interference\cite{hong1987measurement}, of which the former is the kernel unit in a phase-encoding  QKD system and the latter is the core of measurement-device-independent (MDI) QKD. It is worth to be mentioned that we perform the simulation by connecting the fundamental elements according to the structure we exam and then importing the excitation signals. In the simulation, we care about the optical structures rather than the protocols. We compare the simulation results obtained according to the behaviors of the photons and comparing them with the theoretical results to demonstrate that the simulation method is appropriate for QKD system simulation. 

For the convenience of expression in the article, we use a single-photon source as the input excitation signals of the system. The system also supports various types of light sources such as the weak coherent light source and the entanglement light source.

\begin{figure}[ht]
	\centering
	\begin{minipage}[c]{0.45\textwidth}
		\centering
		\includegraphics[width=0.9\textwidth]{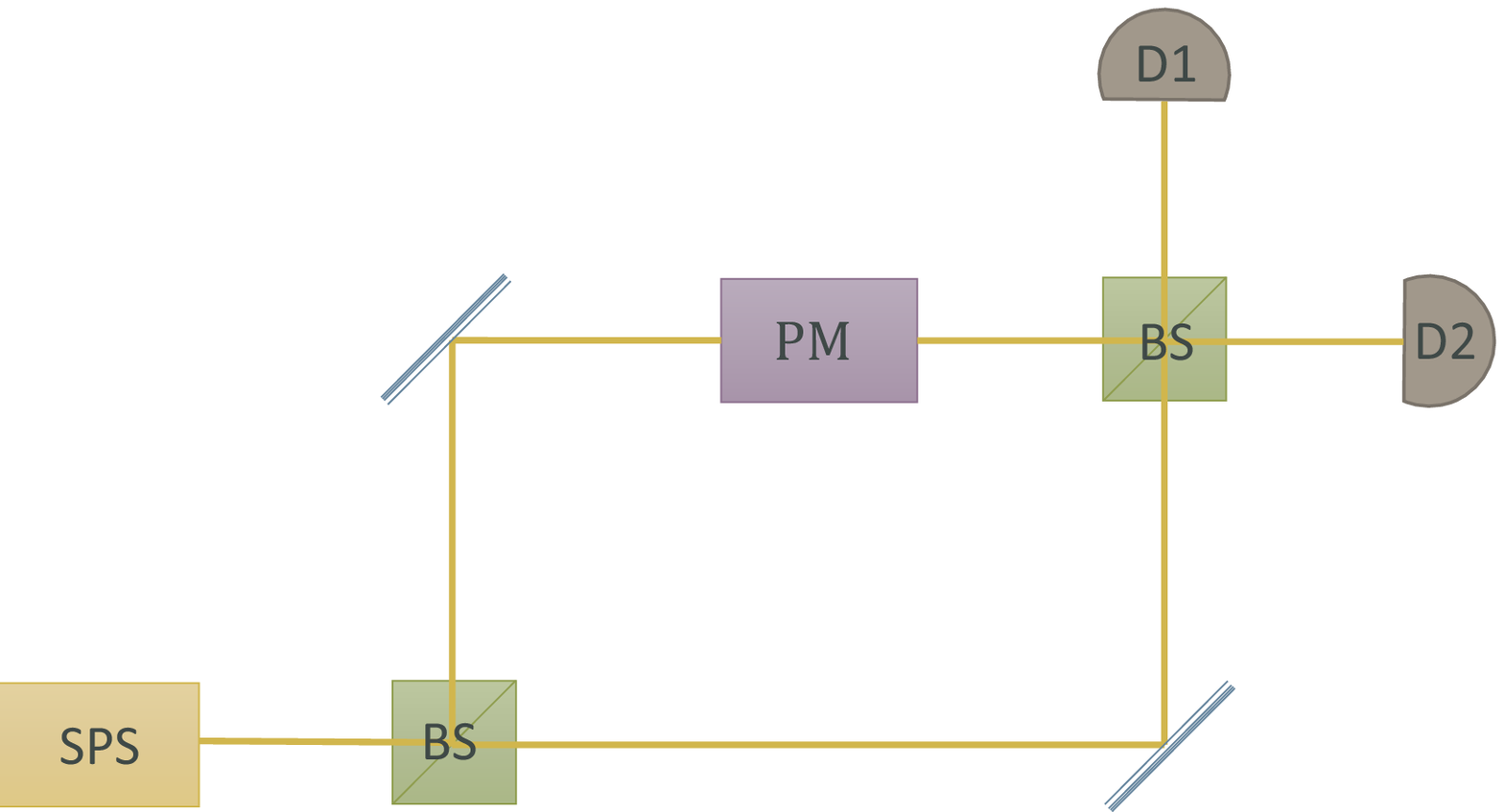}
	\end{minipage}
	\hspace{0.02\textwidth}
	\begin{minipage}[c]{0.45\textwidth}
		\centering
		\includegraphics[width=0.9\textwidth]{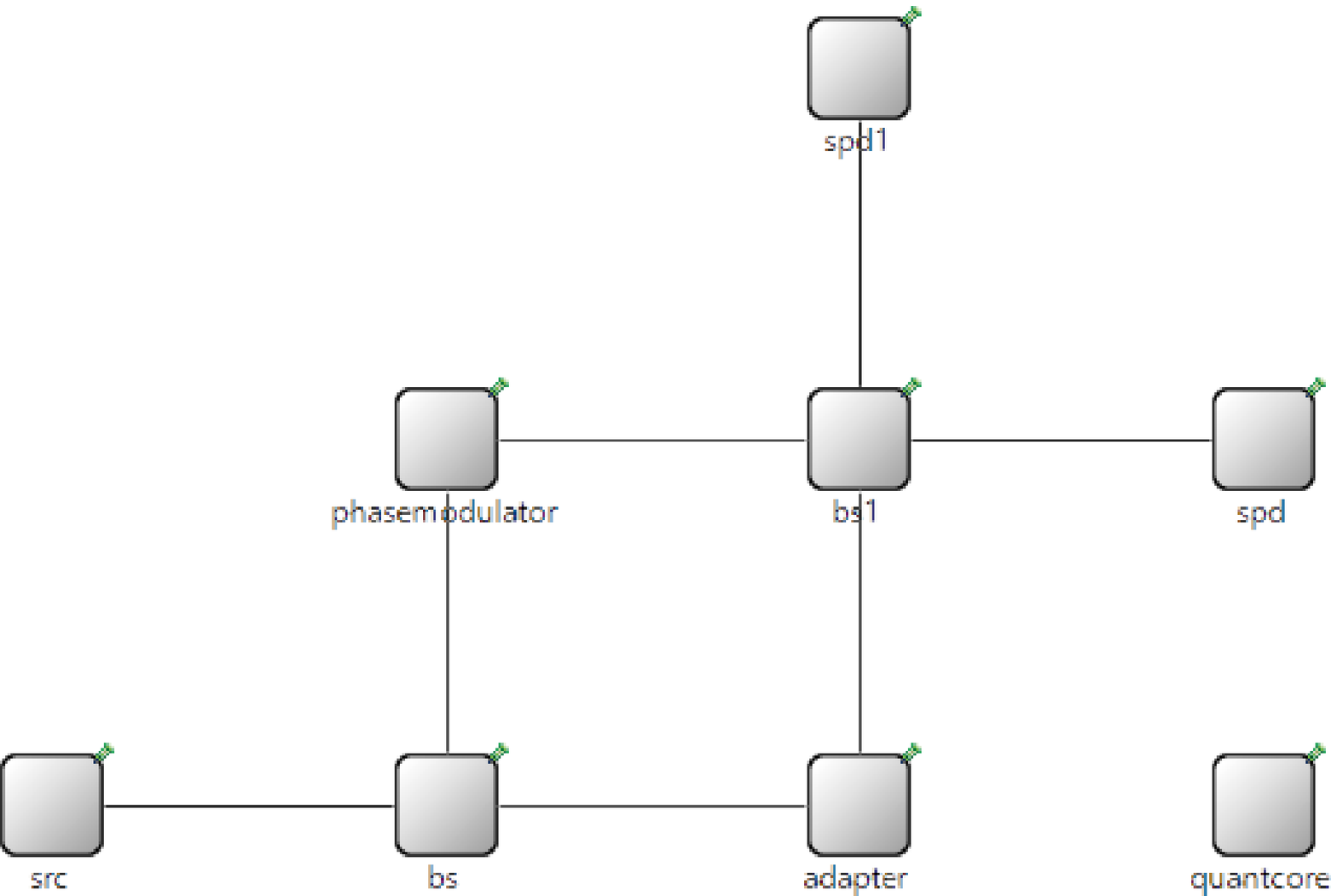}
	\end{minipage}
	\hspace{0.02\textwidth}
	\begin{minipage}[c]{1\textwidth}
		\centering
		\includegraphics[width=0.5\textwidth]{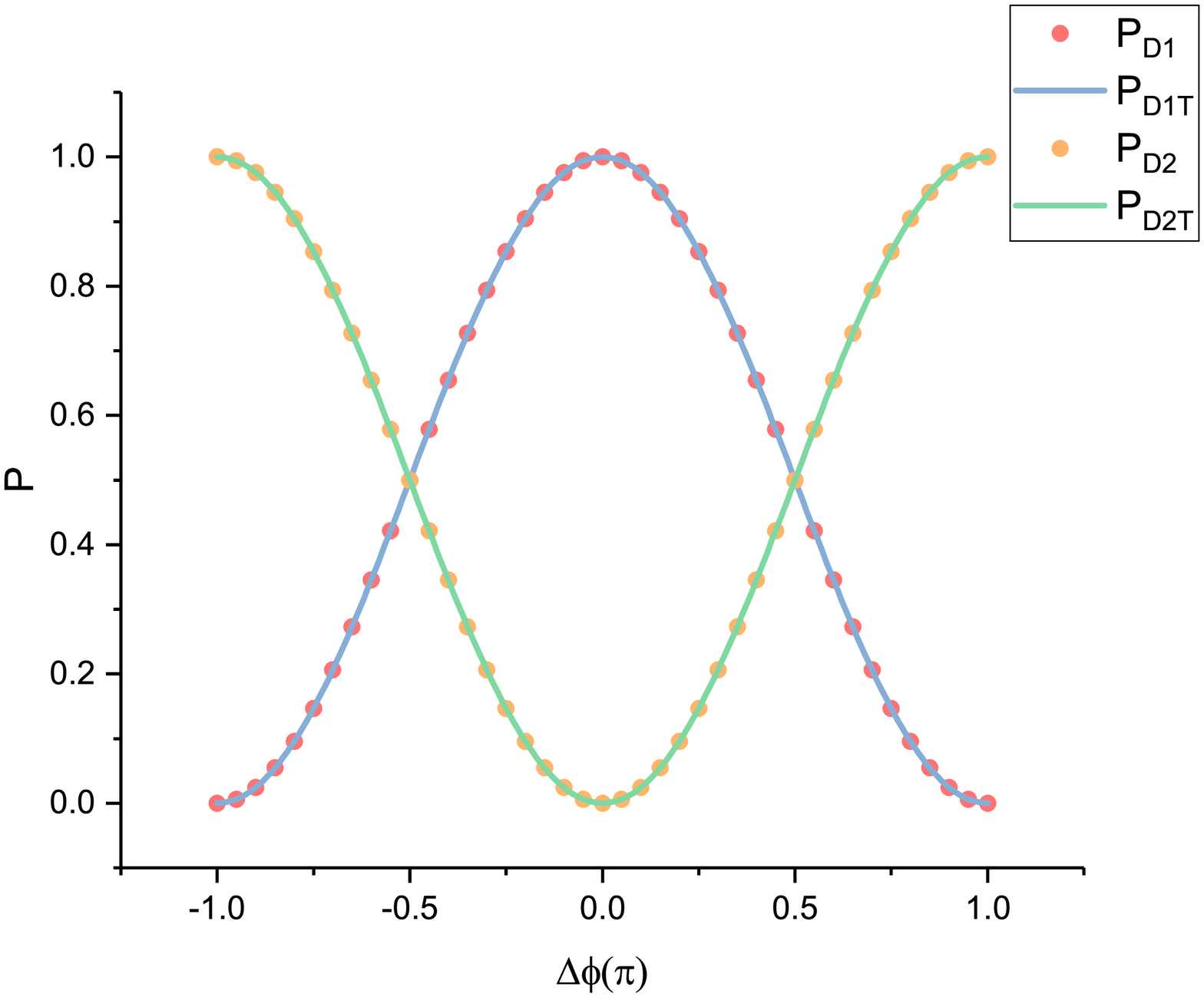}
	\end{minipage}
	\caption{Schematic and simulation network of MZI. \textbf{a}, SPS, single-photon source; BS, beamsplitter; PM, phase modulator; D, detector. \textbf{b}, src, single-photon source; bs, beamsplitter; phasemodulator, phase modulator; adapter, just for aesthetics; spd, detector; quantcore, QuantumCore. \textbf{c}, comparison of simulation results and theoretical calculations. The dotted red and yellow correspond to simulation results $P_{D1}$ and $P_{D2}$, respectively. The solid blue and green are theoretical curves derived from Eq. (\ref{eq27}) respectively.}
	\label{MZI}
\end{figure}

Fig.\ref{MZI} shows the schematic of a MZI, which is widely used in QKD systems, the combination of the basic units in the simulation system, and the simulation results. The evolution of a photon is described as
\begin{equation}
\ket{a}\xrightarrow{\text{BS1}}\frac{1}{\sqrt{2}}\left(\ket{a}+\ket{b}\right)\xrightarrow{\text{PM}}\frac{1}{\sqrt{2}}\left(\ket{a}+e^{i\phi}\ket{b}\right)\xrightarrow{\text{BS2}}\frac{1}{2}\left(\left(e^{i\phi}+1\right)\ket{c}+\left(e^{i\phi}-1\right)\ket{d}\right)
\end{equation}

The theoretical probabilities of detecting the photon at D1 (path $c$) and D2 (path $d$) are given by
\begin{equation}
p_{D1T}=\frac{1+\cos(\phi)}{2}, p_{D2T}=\frac{1-\cos(\phi)}{2}
\label{eq27}
\end{equation}

From Fig.\ref{MZI}(c), we can see that the simulation results are in agreement with the theoretical result, according to Eq.\ref{eq27}. 

\begin{figure}[!t]
	\centering
	\begin{minipage}[c]{0.45\textwidth}
		\centering
		\includegraphics[width=0.6\textwidth]{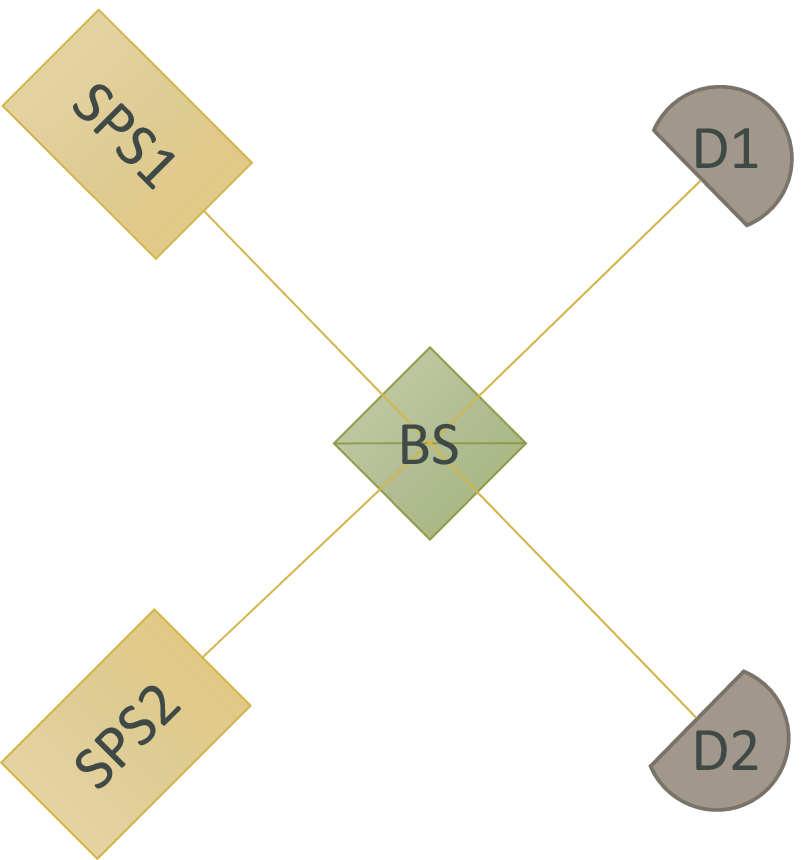}
	\end{minipage}
	\hspace{0.02\textwidth}
	\begin{minipage}[c]{0.45\textwidth}
		\centering
		\includegraphics[width=0.8\textwidth]{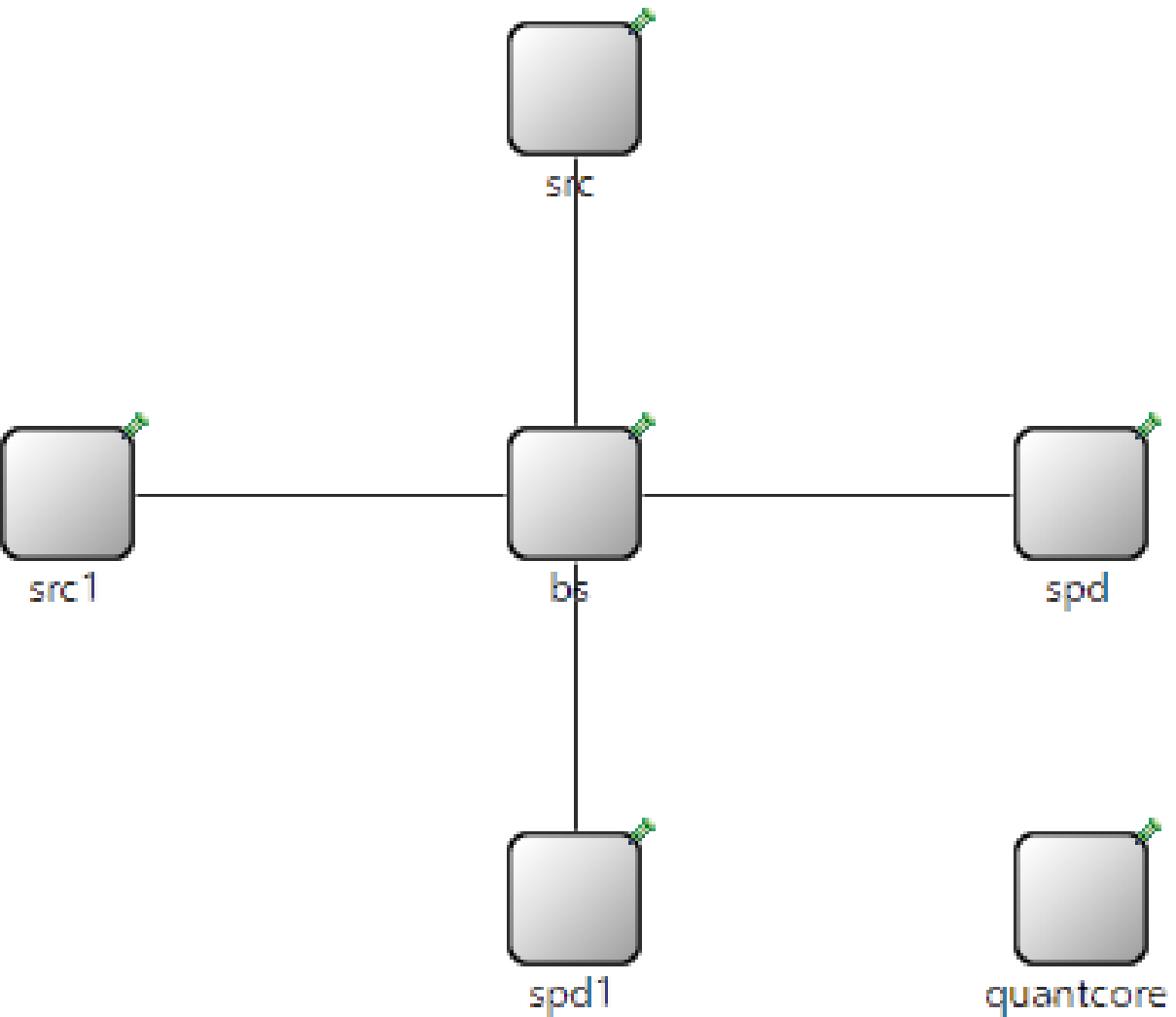}
	\end{minipage}
	\hspace{0.02\textwidth}
	\begin{minipage}[c]{0.45\textwidth}
		\centering
		\includegraphics[width=1\textwidth]{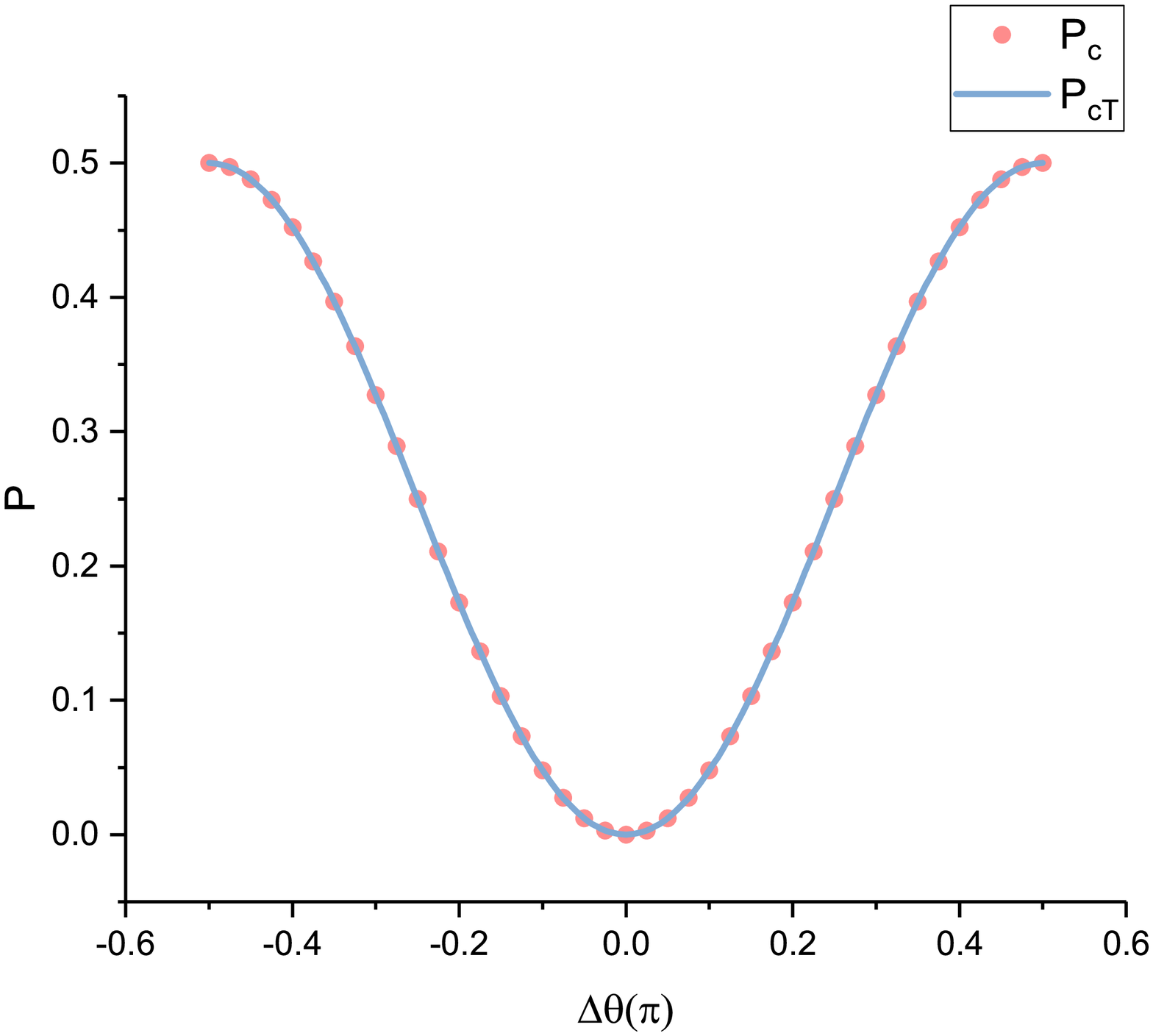}
	\end{minipage}
	\hspace{0.02\textwidth}
	\begin{minipage}[c]{0.45\textwidth}
		\centering
		\includegraphics[width=1\textwidth]{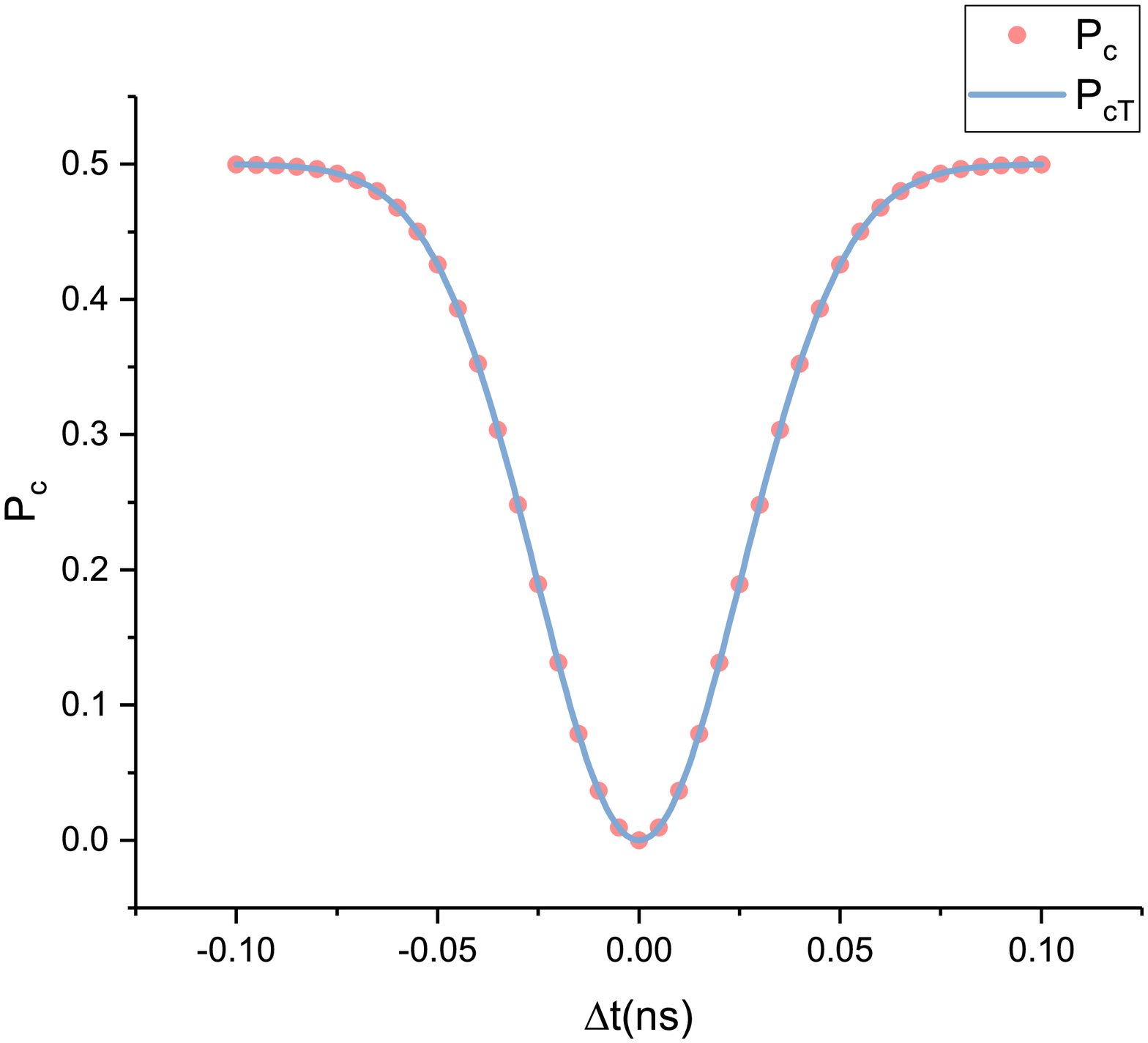}
	\end{minipage}
	\caption{Schematic and simulation network of HOM interferometry. \textbf{a}, SPS, single-photon source; BS, beamsplitter; D, detector. \textbf{b}, src, single-photon source; bs, beamsplitter; spd, detector; quantcore, QuantumCore. \textbf{c}, Comparison of simulation results and theoretical calculations. he dotted red and solid blue correspond to simulation coincidence probability versus $\delta\theta$ and theoretical curves derived from Eq. (\ref{eq30}) respectively. \textbf{d}, The dotted red and solid blue correspond to simulation coincidence probability as a function of $\delta t$ and theoretical curves derived from Eq. (\ref{eq31}) respectively. The $\sigma$ is 65GHz, which is a typical value for a 1550nm laser.}
	\label{HOMsimu}
\end{figure}

The simulation of HOM interference with polarization encoding is shown in Fig.\ref{HOMsimu}. The coincidence probability can be characterized by the relative differences in polarization and arriving time. Without loss of generality, we can depict the relative polarization difference $\delta\theta$ of the two input photon as
$
\ket{\psi_1}=\vu*{a}_{1H}^\dagger\ket{0}, \ket{\psi_2}=\left(\alpha\vu*{b}_{2H}^\dagger+\beta\vu*{b}_{2V}^\dagger\right)\ket{0}
$
where $\alpha=\cos(\delta\theta)$, $\beta=\sin(\delta\theta)$, $\abs{\alpha}^2+\abs{\beta}^2=1$. Then the output state of the BS is given by
\begin{equation}
\ket{\psi_o}=\frac{1}{2}\left(\vu*{a}_{1H}^\dagger+\vu*{b}_{1H}^\dagger\right)\left(\alpha\left(\vu*{a}_{2H}^\dagger-\vu*{b}_{2H}^\dagger\right)+\beta\left(\vu*{a}_{2V}^\dagger-\vu*{b}_{2V}^\dagger\right)\right)\ket{0}
\end{equation}
and the theoretical coincidence probability of getting one photon in each path is given by 
\begin{equation}
p_{cT}=\frac{\abs{\beta}^2}{2}
\label{eq30}
\end{equation}

Taking the arriving time of individual photons into account, the theoretical coincidence probability is \cite{hong1987measurement}
\begin{equation}
p_{cT}=\frac{1}{2}-\frac{1}{2}e^{-\sigma^2\delta t^2}
\label{eq31}
\end{equation}
where $\sigma$ is the standard deviation of the Gaussian frequency distribution, as shown in Eq.\ref{eq2}. The relationship between $\sigma$ and the full width at half maximum (FWHM) of the frequency spectrum is given by
\begin{equation}
\text{FWHM}=2\sqrt{2\ln(2)}\sigma
\label{eq32}
\end{equation}

By comparing the results shown in Fig.\ref{HOMsimu}, we can see that the module-based simulation results of HOM interference are finely consistent with the theoretical results according to Eq.\ref{eq30} and Eq.\ref{eq31}. Since the polarization and the arriving time variations relate to the practical issues of the photon states, the results also indicate the availability of independent operations of the physical variables of photons and show potentials for practical QKD system analysis.

\section{Conclusion and discussion}
We introduced a universal framework for simulation of practical QKD systems. We treat the processing procedure of a QKD system as the cascade transformation to quantum states, which are described with the quantum operators. The processing based on quantum states reflects the non-local properties. The event-driven mechanism of the framework focuses on the detailed quantum procedure of the system, which is commonly ignored in most of the existing simulation works. The multi-dimension descriptions of the signal and the elements make the model universal to evaluate the QKD system of variable protocols, as well as the practical non-idealities of the system. 

Compared with traditional numerical simulations, although our model is at a disadvantage of time consumption, it can precisely and vividly demonstrate the imperfections of practical devices, which is nearly impossible to deduce an analytical formula. Moreover, our program can be further optimized for high running speed. The parallel computing running on a GPU is an option, which can significantly save the time. Also, replacing the Fock basis with the coherent state is convenient for the simulations based on weak coherent sources, that is, the optical elements operate the coherent state directly, although such changes can weaken the universality. Therefore, our model has a distinct advantage of simulating the imperfections of practical devices in quantum language. The time consumption is acceptable and can be reduced further.

It is worth indicating that although this simulation framework provides a possible way for QKD system evaluation, it is constrained when simulating complex systems or signals with multi-photons due to its computational complexity. How to optimize the model and the computational process, as well as to make it valid handle the practical security analysis of the QKD system is, are still challenging works in the future.

\section*{Acknowledgements}
This work has been supported by the National Key Research and Development Program of China (Grant No. 2018YFA0306400), National Natural Science Foundation of China (Grant Nos. 61627820, 61675189, 61622506, 61822115), Anhui Initiative in Quantum Information Technologies(Grant No. AHY030000). We also appreciate Dr. Xuebi An and Yuyang Ding of Anhui Qasky, Co. Ltd. for helpful discussions.

\section*{Appendix A.}

In this section, we show the details regarding the model optical elements Table \ref{tabdetiledcom}. All input and output parameters are denoted by subscript $i$ and $o$, respectively.

\begin{longtable}{ m{50pt}cm{300pt} }
	\caption{Explanation and Output of Optical Element}
	\\\hline
	\toprule
	Optical Element				&Variable Name			&\multicolumn{1}{c}{Explanation}\bigstrut\\ \hline
	\endhead
	\multirow{2}[4]{50pt}{Attenuator}			&Loss				&The loss of attenuator\bigstrut\\ \cline{2-3}
	&\multicolumn{2}{m{320pt}}{$$C_o=C_i\sqrt{10^\frac{-l}{10}}$$ where $l$ is the loss.}\bigstrut\\\hline
	\multirow{2}[4]{50pt}{Bandpass Filter}	&Loss				&The loss of bandpass filter\bigstrut\\ \cline{2-3}
	&\multicolumn{2}{m{320pt}}{$$C_o=C_i\sqrt{10^\frac{-l_\omega}{10}}$$ where $l_\omega$ is the loss when the photon frequency is $\omega$.}\bigstrut\\\hline
	\multirow{2}[4]{50pt}{Circulator}			&Loss				&The insertion loss of circulator\bigstrut\\ \cline{2-3}
	&\multicolumn{2}{m{320pt}}{$$C_o=C_i\sqrt{10^\frac{-l}{10}}$$ where $l$ is the insertion loss.} \bigstrut\\\hline
	\multirow{5}[10]{50pt}{Polarization Modulator}				&Alpha				&The target value of the amplitude of the field in the horizontal\bigstrut\\ \cline{2-3}
	&Beta				&The target value of the amplitude of the field in the vertical\bigstrut\\ \cline{2-3}
	&DeltaPhase		&The target value of the difference between the phase angles of fields in horizontal and vertical directions\bigstrut\\ \cline{2-3}
	&Loss				&The insertion loss of polarization modulator\bigstrut\\ \cline{2-3}
	&\multicolumn{2}{m{320pt}}{$$\alpha_o=\alpha, \beta_o=\beta, \delta\theta_o=\delta\theta, C_o=C_i\sqrt{10^\frac{-l}{10}}$$ where $\alpha, \beta, \delta\theta$ are the target value of the polarization parameters and $l$ is the insertion loss.}\bigstrut\\\hline
	\multirow{5}[10]{50pt}{Phase Modulator}				&Phase				&The target value of the phase\bigstrut\\ \cline{2-3}
	&Loss				&The insertion loss of phase modulator\bigstrut\\ \cline{2-3}
	&\multicolumn{2}{m{320pt}}{$$\varphi_o=\varphi, C_o=C_i\sqrt{10^\frac{-l}{10}}$$ where $\varphi$ are the target value of phase and $l$ is the insertion loss.}\bigstrut\\\hline
	\multirow{3}[4]{50pt}{Isolator}				&Loss						&The insertion loss of isolator\bigstrut\\ \cline{2-3}
	&IsolationLoss			&The isolation loss of isolator\bigstrut\\ \cline{2-3}
	&\multicolumn{2}{m{320pt}}{$$\begin{cases}C_o=C_i\sqrt{10^\frac{-l+l_is}{10}},&\text{transmitting along the forward direction}\\C_o=C_i\sqrt{10^\frac{-l}{10}},&\text{transmitting along the reverse direction}\end{cases}$$where $l$ is the insertion loss and $l_is$ is the isolation loss.}\bigstrut\\\hline
	\multirow{2}[4]{50pt}{1x2 Optical Switch}	&Loss						&The insertion loss of optical switch\bigstrut\\ \cline{2-3}
	&IsolationLoss			&The isolation loss of optical switch\bigstrut\\ \cline{2-3}
	&\multicolumn{2}{m{320pt}}{$$\begin{cases}C_o=C_i\sqrt{10^\frac{-l+l_is}{10}},&\text{emitting from the desired output port}\\C_o=C_i\sqrt{10^\frac{-l}{10}},&\text{emitting from the undesired output port}\end{cases}$$where $l$ is the insertion loss and $l_is$ is the isolation loss.}\bigstrut\\\hline
	\multirow{4}[8]{50pt}{Waveplate}			&RelativePhase	&The phase shift between polarization components, $\pi$ for a half-wave plate and $\pi/2$ for a quarter-wave plate\bigstrut\\ \cline{2-3}
	&OffsetAngle		&The angle of the fast axis\bigstrut\\ \cline{2-3}
	&Loss				&The insertion loss of the waveplate\bigstrut\\ \cline{2-3}
	&\multicolumn{2}{m{380pt}}{
		The Stokes vector of input photon state is given by $$S_i=\left(1, \alpha_i^2-\beta_i^2, 2\alpha_i\beta_i\cos{\delta\theta_i}, 2\alpha_i\beta_i\sin{\delta\theta_i}\right)^T$$
		where $\alpha$ and $\beta$ are the amplitudes of the field in the horizontal and vertical direction respectively, 
		$\delta\theta$ is the difference between the phase angles of fields in horizontal and vertical directions.
		
		The Mueller matrice of the waveplate is given by $$M=\begin{pmatrix}\begin{smallmatrix}  1&0&0&0 \\  0&\cos[2](2\theta)+\cos(\delta)\sin[2](2\theta)&\cos(2\theta)\sin(2\theta)-\cos(2\theta)\cos(\delta)\sin(2\theta)&\sin(2\theta)\sin(\delta) \\  0&\cos(2\theta)\sin(2\theta)-\cos(2\theta)\cos(\delta)\sin(2\theta)&\cos(\delta)\cos[2](2\theta)+\sin[2](2\theta)&-\cos(2\theta)\sin(\delta) \\  0&-\sin(2\theta)\sin(\delta)&\cos(2\theta)\sin(\delta)&\cos(\delta)\end{smallmatrix}\end{pmatrix}$$
		where $\theta$ is the OffsetAngle and $\delta$ is the RelativePhase.
		Then the Stokes vectors of output photon state is given by $$S_o=S_iM=\left(S_0, S_1, S_2, S_3\right)^T$$
		So the polarization parameters of output photon state become $$\alpha_o=\sqrt{1+S_1}, \beta_o=\sqrt{1-S_1}, \delta\theta_o=
		\begin{cases}
		0&\alpha_o\beta_o=0\\
		\arccos{\left(\frac{S_2}{2\alpha_o\beta_o}\right)}&\alpha_o\beta_o\neq0, S_2\geq0\\
		-\arccos{\left(\frac{S_2}{2\alpha_o\beta_o}\right)}&\alpha_o\beta_o\neq0, S_2<0
		\end{cases}$$
		Also, the output normalization coefficient is given by
		$$C_o=C_i\sqrt{10^\frac{-l}{10}}$$
		where $l$ is the insertion loss.
	}\bigstrut\\\hline
	\multirow{3}[4]{50pt}{Single Mode (SM) Fiber}				&Alpha		&The loss of SM fiber per kilometer\bigstrut\\ \cline{2-3}
	&Length	&The length of SM fiber\bigstrut\\ \cline{2-3}
	&Sigma	&The variance of the normal distribution for random disturbance\bigstrut\\ \cline{2-3}
	&Expectation	&The expectation of the normal distribution for random disturbance\bigstrut\\ \cline{2-3}
	&\multicolumn{2}{m{320pt}}{$$C_o=C_i\sqrt{10^\frac{-Al}{10}}, a_o=a_i+\delta, \delta\sim N(\mu_a,\sigma_a)$$ where $A$ is the Alpha, $l$ is the Length,  $\mu_a$ and $\sigma_a$ are the Expectation and Sigma, respectively, and $a\in\{\varphi, \alpha, \beta, \theta\}$. For an ideal polarization maintaining fiber, the polarization-dependent $\mu$ and $\sigma$ should be $0$.}\bigstrut\\\hline
	\multirow{3}[4]{50pt}{Faraday Mirror (FM)}	&Loss						&The insertion loss of FM\bigstrut\\ \cline{2-3}
	&Theta					&The faraday rotation of a single pass\bigstrut\\ \cline{2-3}
	&\multicolumn{2}{m{380pt}}{A Faraday mirror is a combination of a Faraday rotator and an ordinary mirror whose Jones matrix is given
		by\cite{mo2005faraday} $$FM=
		\begin{pmatrix}
		\cos(\theta)	&\sin(\theta) \\
		-\sin(\theta)	&\cos(\theta)
		\end{pmatrix}
		\begin{pmatrix}
		1	&0 \\
		0	&-1
		\end{pmatrix}
		\begin{pmatrix}
		\cos(\theta)	&-\sin(\theta) \\
		\sin(\theta)	&\cos(\theta)
		\end{pmatrix}$$
		therefore,
		$$FM=\begin{pmatrix}
		\cos[2](\theta)-\sin[2](\theta)	&-2\sin(\theta)\cos(\theta) \\
		-2\sin(\theta)\cos(\theta)		&\sin[2](\theta)-\cos[2](\theta)
		\end{pmatrix}
		$$
		where $\theta$ is the Theta. Then the output photon state is given by $$e^{i\varphi}\begin{pmatrix}\alpha_o \\\beta_oe^{i\theta_o}\end{pmatrix}=FM\begin{pmatrix}\alpha_i \\\beta_ie^{i\theta_i}\end{pmatrix}, \varphi_o=\varphi_i+\varphi$$ Also, the output normalization coefficient is given by $$C_o=C_i\sqrt{10^\frac{-l}{10}}$$ where $l$ is the insertion loss.}\bigstrut\\\bottomrule
		\label{tabdetiledcom}
\end{longtable}

\bibliography{reference}
\end{document}